\documentclass[journal]{IEEEtran}

\IEEEoverridecommandlockouts

\usepackage{graphicx}
\usepackage{setspace} % put this in preamble

\usepackage{amsmath}
\usepackage{amssymb}
\usepackage{graphicx}
\usepackage{cite}
\usepackage{amsmath}
\usepackage{amssymb}
\usepackage{amsthm}
\usepackage{lipsum}
\usepackage{xcolor}
\usepackage[0]{editing}
\usepackage{algpseudocode}
\usepackage{multirow}
\usepackage{arydshln}
\usepackage{bm}
\usepackage{subfig}
\usepackage{stfloats}% <-- added
\usepackage[linesnumbered,ruled,vlined]{algorithm2e}

\theoremstyle{definition}

\usepackage[linesnumbered,ruled,vlined]{algorithm2e}

\def\BibTeX{{\rm B\kern-.05em{\sc i\kern-.025em b}\kern-.08em
    T\kern-.1667em\lower.7ex\hbox{E}\kern-.125emX}}

\begin{document}

\title{Channel-Correlation-Based Access Point Selection and Pilot Power Allocation for Cell-Free Massive MIMO }

\author{Saeed~Mohammadzadeh,~\IEEEmembership{Member,~IEEE,}  %$^{\orcidlink{0000-0002-8793-2577}}$ 
    Rodrigo~C.~De~Lamare,~\IEEEmembership{Fellow~Member,~IEEE,}\\
    Kanapathippillai~Cumanan,~\IEEEmembership{Senior~Member,~IEEE,}
    and~Hien~Quoc~Ngo,~\IEEEmembership{Fellow,~IEEE.} \vspace{-0.75em}
    \thanks{S. Mohammadzadeh, R. C. De Lamare, and K. Cumanan are with the School of Physics, Engineering and Technology, University of York. UK. (email: Saeed.mzadeh@ieee.org; rodrigo.delamare@york.ac.uk; kanapathippillai.cumanan@york.ac.uk).}
  \thanks{H. N. Ngo is with the School of Electronics, Electrical Engineering and Computer Science, Queen's University Belfast, (email:hien.ngo@qub.ac.uk).}% <-this % stops a space

% \thanks{The work of S. Mohammadzadeh and K. Cumanan were supported by the UK Engineering and Physical Sciences Research Council (EPSRC) under grant number EP/X01309X/1.}
% \thanks{The work of O. A. Dobre was supported in part by the Canada Research Chairs Program CRC-2022-00187 and the NSERC Discovery grant RGPIN-2019-04123.}
}

\maketitle
\begin{abstract}
This paper proposes a dynamic access point (AP) selection and pilot power allocation (DAPPA) framework for uplink cell-free massive multiple-input multiple-output (CFmMIMO) systems, aiming to mitigate inter-user interference and improve overall spectral efficiency (SE). A hierarchical correlation-based clustering algorithm is developed to group APs according to their channel correlation, enabling each user to be associated with APs that simultaneously provide strong channel gains and low mutual correlation. This association ensures reliable connectivity, maximizes coherent combining gains, and reduces inter-user interference, while also allowing the number of AP clusters to be adjusted flexibly, without the need to reorganize the network completely. By maintaining links to low-correlated APs, the proposed scheme reduces the need for frequent channel state information (CSI) estimation and minimizes network-wide update overhead. To enhance scalability, a user-capacity constraint per AP is incorporated, preventing hardware overload and alleviating the effects of pilot reuse. Furthermore, an effective pilot power allocation strategy is introduced to boost the signal-to-interference-plus-noise ratio (SINR) during channel training. This is formulated as a weighted sum-rate maximization (WSRM) problem and solved iteratively using a quadratic transform, which enables efficient optimization while ensuring fairness and high-quality service across all users. Numerical results demonstrate that the proposed method delivers significant SE gains, maintains performance in high-density multi-user scenarios, and converges faster than benchmark schemes. 
\end{abstract}

\IEEEpeerreviewmaketitle

\section{Introduction}

Cell-free massive multiple-input multiple-output (CFmMIMO) systems are a promising technology for future wireless communication networks, offering seamless connectivity and superior service quality to users \cite{ngo2015cell,ngo2017cell}. In CFmMIMO networks, a large number of distributed access points (APs) collaboratively serve user devices across a wide area without the need for traditional cell boundaries. These APs connect to one or several central processing units (CPUs) via fronthaul links, with the CPUs interconnected through a backhaul network \cite{bashar2019max,demir2021foundations,bashar2019uplink}. CFmMIMO networks offer enhanced spectral and energy efficiency, leveraging benefits like favorable propagation, high macro-diversity, efficient resource allocation, and channel hardening \cite{bashar2019energy,hao2023user,chong2024performance}. 

\subsection{Prior and Related Works}
\indent In standard CFmMIMO networks, all APs serve all the users simultaneously \cite{bjornson2019making}. However, in these systems, the required number of channel measurements increases linearly with both the number of APs and users, resulting in substantial channel state information (CSI) signaling overhead and higher signal processing complexity \cite{bjornson2020scalable}. In reality, a user's signal power is primarily influenced by a limited number of nearby APs, as path loss increases exponentially with distance. This suggests that coordinating all APs to serve every user jointly is unnecessary. To overcome these challenges, recent approaches assign tailored AP clusters to users, enhancing the scalability of CFmMIMO networks \cite{dai2014uplink,lin2017graph,ngo2017total,buzzi2017cell,ammar2019power,lin2019user,alonzo2019energy,chen2020structured,wei2022user,sheikh2022capacity,li2023access,wang2023clustered, mohammadzadeh2025association}.

In \cite{dai2014uplink}, a user-centric overlapped clustering algorithm based on a graph-theoretical framework is proposed. The AP selection in \cite{ngo2017total} is proposed based on two schemes, utilizing the received power-based selection and the largest-scale fading (LSF)-based approach \cite{shakya2020joint,duan2022pilot}. In \cite{buzzi2017cell}, a method was introduced that utilizes a dynamic set of APs to serve each user based on the virtual cell approach and system performance, or large-scale fading \cite{ammar2019power}. The authors in \cite{lin2019user} propose a method in which each AP serves the users within its coverage of some radius. The work in \cite{chen2020structured} proposed a structured massive access-based pilot assignment strategy that forms groups with users having a minimum intersection of shared serving APs and assigns orthogonal pilots. In \cite{wei2022user}, the AP selection problem was introduced as a local altruistic game based on its inherent interference relationship. The distance-based AP selection and random-access point selection algorithms are used to choose APs in \cite{sheikh2022capacity}. 

In \cite{wang2023clustered}, a clustered cell-free network architecture was proposed, where the entire network is divided into non-overlapping sub-networks, each operating independently with cooperative transmission. In \cite{li2023access}, an AP selection strategy based on quantum bacterial foraging optimization is proposed, which encodes the connection relationship between APs and users in the form of qubits. \add{Authors in \cite{huang2023joint} propose a joint pilot assignment scheme based on user clustering and graph coloring, where AP selection is first used to group users, followed by graph-based pilot allocation.} The scheme in \cite{yemini2019virtual} utilizes a hierarchical clustering algorithm employing the Euclidean distance between the APs to group APs based on the minimax linkage criterion \cite{bien2011hierarchical}, while the deep learning-based version is introduced in \cite{tan2024energy}. In \cite{wang2021performance}, user clustering is performed using a hierarchical method, while AP selection is based on large-scale fading decoding and the normalized channel estimation mean square error. In \cite{banerjee2023access}, the authors developed a multi-agent reinforcement learning (MARL) algorithm for AP selection and clustering. Each AP is an agent in the MARL algorithm, trained to select which users to serve near-optimally. Recently, clustering based on the sum-rate criterion was reported in \cite{mashdour2024clustering} and showed enhanced AP clustering.

\indent Moreover, it is well known that to achieve an efficient resource allocation in wireless systems, power control (PC) is crucial, helping to reduce interference \cite{mosayebi2020linear,attarifar2020subset,xiao2023pilot}, boost performance, and meet quality of service requirements for users in CFmMIMO networks \cite{bjornson2014massive,zhao2020power}. By optimizing both pilot and data power, spatial diversity in the wireless channel can be better exploited, reducing the impact of pilot contamination, inter-user interference, and noise \cite{van2018large, mohammadzadeh2025pilot}. However, designing efficient algorithms to manage PC in large-scale CFmMIMO networks remains a complex task, largely due to the intricate nature of the optimization problem and practical constraints such as computational costs and signaling overhead. The works in \cite{cheng2016optimal,ghazanfari2018optimized,van2020power, zhang2021deep, sarker2023pilot} explored the impact of pilot and PC in the uplink of cell-based systems. However, the optimization algorithms used in these cellular systems are not directly applicable to cell-free systems. In cell-free systems, each serving AP must estimate the channels to its users using the uplink pilot signal. Moreover, prior work on cell-free systems focused on optimizing data power \cite{ngo2017cell}, \cite{nguyen2018optimal} while assuming either full or fixed power for transmitting pilot signals.

Using a bisection search algorithm, the study in \cite{ngo2017cell} addressed the max-min PC problem in CFmMIMO networks. In \cite{mai2018pilot}, an algorithm was introduced using a pilot PC scheme to minimize the maximum channel estimation error among users. It employed the data power allocation strategy from \cite{ngo2017cell} to address a max-min fairness problem, while AP selection was based on the LSF criteria. However, the key challenge with this approach lies in the time-consuming process of finding the maximum LSF values for all users across all APs. In \cite{masoumi2018joint}, a joint PC method was considered for single-antenna APs, where users have limited total energy. The work in \cite{mai2020design} explored a first-order Taylor approximation approach, jointly optimizing pilot and data power using a greedy power algorithm. In user-centric scenarios \cite{braga2021joint}, authors proposed a joint pilot and data PC by combining successive convex approximation and geometric programming (GP). Furthermore, the work in \cite{liu2022joint} employed a sequential convex approximation algorithm to solve a GP problem for jointly controlling pilot and data power. Moreover, a fractional PC policy for uplink CFmMIMO was presented in \cite{nikbakht2019uplink}, which can be fully distributed and uses large-scale quantities. Inspired by weighted minimum mean square error (WMMSE) and fractional programming, the algorithm in \cite{chakraborty2020efficient} proposed a fair power allocation. Deep neural networks (DNNs) were introduced in \cite{bashar2020exploiting,zaher2022learning,salaun2022gnn} to reduce complexity, including convolutional neural network models, distributed DNNs, and graph neural networks for efficient PC optimization with maximum ratio transmission beamforming. 
\subsection{Contribution} 
Motivated by the above discussion, we propose a dynamic AP selection and pilot power allocation schemes to enhance the spectral efficiency (SE) of uplink CFmMIMO systems in this work. We form AP clusters instead of user clusters and then associate users with these clustered APs based on channel correlation. 
Using a specified linkage criterion, the algorithm builds a dendrogram that shows how clusters merge or split as their number changes. Its key advantage is that adjusting the number of clusters only affects the involved clusters, leaving others unchanged. In contrast, methods like K-means or spectral clustering require reclustering the entire network when the cluster count changes, leading to widespread CSI re-estimation, an undesirable overhead in wireless networks. To address this, we propose an algorithm that efficiently adapts the number of clusters to the network’s current state without requiring a full-scale update. Additionally, our approach relies only on local CSI, which is used for user association, making it a scalable and efficient solution. The primary advantage of the proposed method over traditional techniques lies in its correlation-based clustering approach. This method is designed to maximize beamforming gain while effectively minimizing interference. 

\indent To enhance clustered-AP network performance, we propose optimizing pilot power allocation to improve users’ SINR during training. The system dynamically adjusts pilot powers to maximize the weighted sum-rate under power constraints, formulated as a WSRM problem solved via quadratic transform. This ensures consistent, high-quality service, with numerical results showing superior spectral efficiency compared to benchmarks. The key contributions and findings are summarized as follows:
\begin{itemize}
    \item The proposed scheme enhances uplink CFmMIMO SE by dynamically clustering APs and allocating pilot power. AP clusters are formed using a hierarchical correlation-based method, associating users with APs that offer both strong channel gains and low correlation to ensure reliable connectivity and reduced interference.
    
    \item The method allows flexible cluster adjustments without full network reclustering, reducing CSI re-estimation and minimizing network-wide overhead.
    
    \item User–AP associations maximize coherent combining gains and reduce interference by selecting APs with strong, low-correlated channels, achieving balanced service quality (though at the cost of higher backhaul demand).
    
    \item Scalability is supported by limiting the number of users per AP, preventing overloads due to hardware limits or pilot reuse, and sustaining performance in dense networks.
    
    \item Pilot power allocation is optimized to improve SINR during training, formulated as a WSRM problem and solved for consistent, high-quality service.
\end{itemize}

\subsection{Outline}
The remainder of the paper is organized as follows. Section II describes the system model. Section III introduces the proposed algorithm for AP selection and pilot power allocation, aimed at optimizing AP clustering and improving SE. This section also provides a summary of the algorithms and an analysis of their computational complexity. Section IV presents simulation results, demonstrating the advantages of the proposed approach. Finally, Section V concludes the paper.

\subsection{Notations} We represent vectors using bold lowercase letters and matrices using bold uppercase letters. $\mathbb{E} \{\cdot\}$ stands for the statistical expectation of random variables, and a circular symmetric complex Gaussian matrix with covariance $\mathbf{Z}$ is denoted by $\mathcal{CN}(0,\mathbf{Z})$. The diagonal matrix operator is denoted by $\mathrm{diag}(\cdot)$. 
The sets of complex and real numbers are represented by $\mathbb{C}$ and $\mathbb{R}$, respectively. 
The notations $(\cdot)^\mathrm{T}$, $(\cdot)^\ast$, $(\cdot)^{-1}$, and $(\cdot)^\mathrm{H}$ denote the transpose, conjugate, inverse, and conjugate transpose (Hermitian), respectively. 
The Euclidean norm and absolute value are written as $\|\cdot\|$ and $|\cdot|$, respectively. 
Superscripts $^\mathrm{p}$ and $^\mathrm{d}$ are used to indicate variables or parameters related to pilot and data transmission, respectively. 
The notation $\left(\cdot\right)^{\downarrow}$ denotes descending order, and $\mathbf{I}$ represents the identity matrix.

\begin{figure}[!]
	\centering 
	\includegraphics[width=3in,height=1.5in]{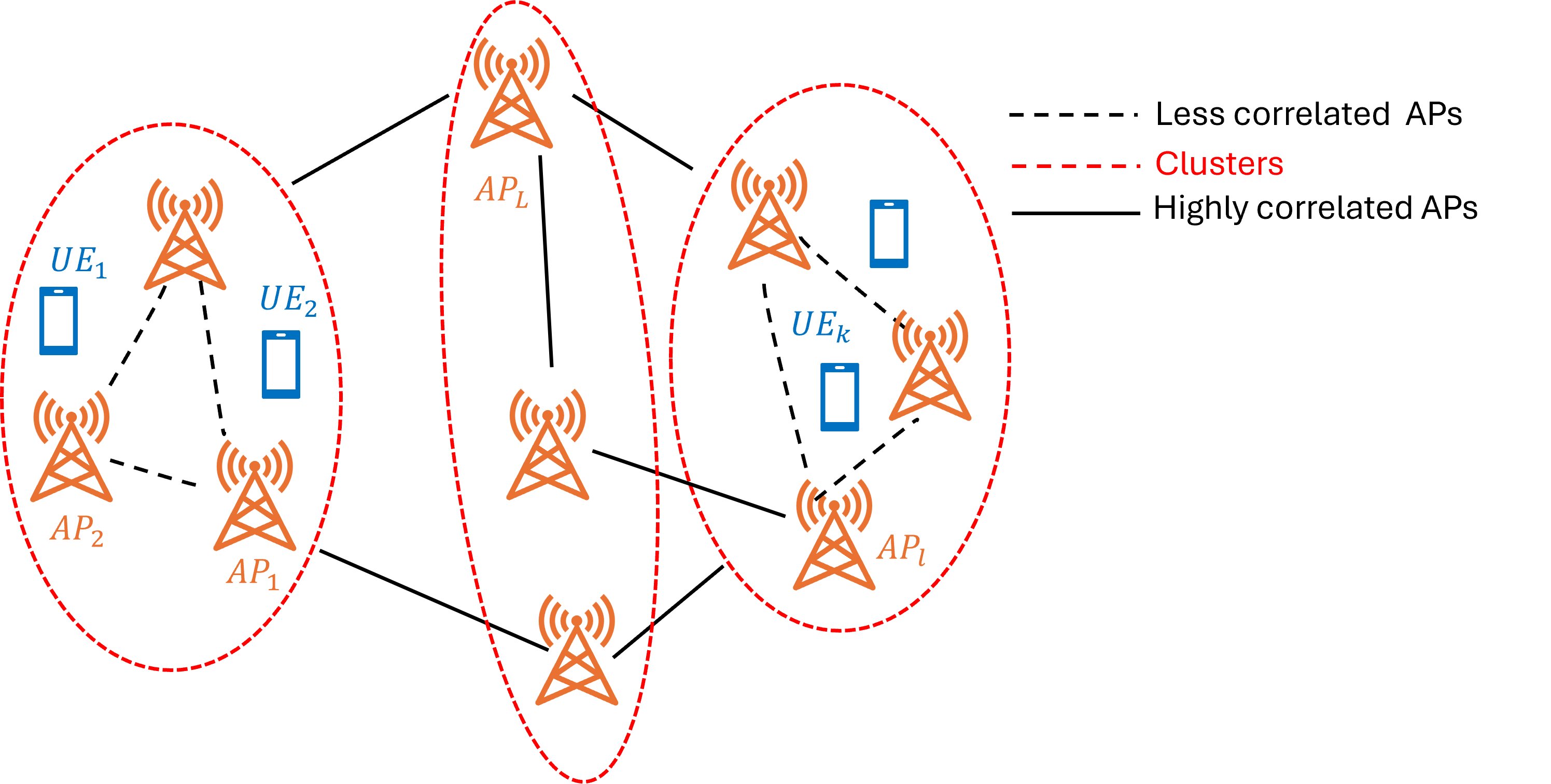}
	\caption{AP clustering for CFmMIMO system.}\label{CF figure}
\end{figure} 

\section{System Model}

We consider a scalable CFmMIMO system as shown in Fig.~\ref{CF figure} with $L$ randomly distributed APs that serve $U$ single-antenna UEs; a reasonable assumption should be $U \ll L$. All APs are equipped with $M$ antennas and connected to a CPU via fronthaul links. The system operates in time-division duplex (TDD) mode, and we assume that all APs and users are perfectly synchronized. For each user $u$, a subset of APs denoted by $\mathcal{A}_u$ is selected to provide service as described in \ref{AP clustering}. We consider a system that operates in a block-fading channel, where the time and frequency plane is divided into coherence blocks, $\tau_c$, in which $\tau$ and $\tau_c-\tau$ denote the uplink pilot and data transmission durations (in symbols), respectively. The channel between the $l^\text{th}$ AP and the $u^\text{th}$ user is denoted by the channel $\mathbf{h}_{lu} \in \mathbb{C}^{M \times 1}$. In each block, an independent realization from a correlated Rayleigh fading distribution is drawn as $\mathbf{h}_{lu} \sim \mathcal{CN}(0, \mathbf{B}_{lu})$ where $\beta_{lu} = \text{tr}(\mathbf{B}_{lu})/M$ is the large-scale fading that describes geometric pathloss and shadowing, and $\mathbf{B}_{lu} = \mathbb{E} \{  \hat{\mathbf{h}}_{lu} \hat{\mathbf{h}}_{lu}^\mathrm{H} \}$.

\subsection{Uplink Training and Channel Estimation}
We consider $\tau$ mutually orthogonal pilot sequences $\{\boldsymbol{\phi}_1, \ldots, \boldsymbol{\phi}_{\tau}\}$ of length $\tau$, such that $\|\boldsymbol{\phi}_t\|^2 = \tau$. These pilots are assigned to UEs in a deterministic but otherwise arbitrary manner. Since in practical networks $U>\tau$, multiple UEs must share the same pilot. The pilot index assigned to UE $u$ is denoted by $t_u \in \{1,\ldots,\tau\}$, and we define $\mathcal{P}_u \subseteq \{1,\ldots,U\}$ as the set of UEs reusing pilot $t_u$, including UE $u$.  

During the uplink training phase, the received pilot signal at AP $l$, denoted by $\mathbf{Y}_l \in \mathbb{C}^{M \times \tau}$, is given by
\begin{align}
    \mathbf{Y}_l = \sum_{i=1}^U \sqrt{p_i^\mathrm{p}}\,\mathbf{h}_{li}\boldsymbol{\phi}_{t_i}^\mathrm{T} + \mathbf{N}_l,
\end{align}
where $p_i^\mathrm{p} $ is the transmit power of UE $i$, and $\mathbf{N}_l \in \mathbb{C}^{M \times \tau}$ is the receiver noise matrix with i.i.d.\ entries distributed as $\mathcal{N}_\mathbb{C}(0,\sigma^2)$.
To estimate the channel of UE $i$, AP $l$ correlates $\mathbf{Y}_l$ with the normalized pilot $\boldsymbol{\phi}_{t_u}/\sqrt{\tau}$, yielding
\begin{align} \label{ylp}
    \mathbf{r}_{l}^\mathrm{p} 
    &= \frac{1}{\sqrt{\tau}} \mathbf{Y}_l \boldsymbol{\phi}_{t_u}^* = \sum_{i \in \mathcal{P}_u} \sqrt{p_i^\mathrm{p} \tau}\,\mathbf{h}_{li} + \mathbf{n}_{l}^{p},
\end{align}
where $\mathbf{n}_l^{p} = \mathbf{N}_l \boldsymbol{\phi}_{t_u}^*/\sqrt{\tau} \sim \mathcal{N}_\mathbb{C}(\mathbf{0},\sigma^2\mathbf{I}_N)$.
Using standard MMSE estimation theory \cite{bjornson2017massive}, the channel estimate of $\mathbf{h}_{lu}$ is
\begin{align} 
\hat { \mathbf {h}}_{lu} =  \sqrt{p_u^\mathrm{p} \tau}\,\mathbf{B}_{lu}\boldsymbol{\Psi}_{l_{t_u}}^{-1}\mathbf{r}_{l}^p, 
\end{align}
where $\boldsymbol{\Psi}_{l_{t_u}} = \mathbb{E} \{ \mathbf{r}_l^\mathrm{p} (\mathbf{r}_l^\mathrm{p})^\mathrm{H}  \} = \sum_{i \in \mathcal{P}_u} \tau p_i^\mathrm{p} \mathbf{B}_{li} + \sigma^2\mathbf{I}_M$. The estimate and error are independent Gaussian vectors with
\begin{align}
    \hat{\mathbf{h}}_{lu} &\sim \mathcal{N}_\mathbb{C}\!\left(\mathbf{0},\, p_u^\mathrm{p} \tau \mathbf{B}_{lu} \boldsymbol{\Psi}_{l_{t_u}}^{-1} \mathbf{B}_{lu}\right).
    \end{align}
Since multiple UEs reuse the same pilot, the correlated term in \eqref{ylp} introduces interference, leading to the well-known \emph{pilot contamination} effect that degrades channel estimation accuracy, similar to conventional cellular mMIMO systems.

\indent To ensure system scalability, we define a matrix $\mathbf{A} \in \mathbb{R}^{L \times U}$ that specifies the AP-UE associations, where $\mathbf{A}_{lu} = 1$ indicates that user $u$ is served by AP $l$, and $\mathbf{A}_{lu} = 0$ otherwise. For clarity we define $\mathcal{A}_u = \{ l : \mathbf{A}_{lu} = 1, l \in \{1, \cdots, L\} \}$ as the subset of APs assigned to serve user $u$.

\subsection{Uplink Data Transmission and Achievable SE}
During the uplink data transmission, all UEs send their data to the APs using the same time-frequency resource. The received signal, $\mathbf {r}_l^{\mathrm {d}} \in \mathbb{C}^{M \times 1}$, at the $l^\text{th}$ AP is given by
\begin{equation} 
\mathbf {r}_l^{\mathrm {d}} = \sum _{i=1}^{U} \sqrt{{p_{u}^\mathrm {d}}} \mathbf {h}_{li}s_{i} + \mathbf {n}_{l}^{\mathrm {d}}, 
\end{equation}
where ${p_{i}^\mathrm {d}}$ is the transmit data power of user $i$,  $s_i$ with $\mathbb{E} \{ | s_i |^2 | \}=1$, is the data symbol transmitted by $u^\text{th}$ user, and $\mathbf {n}_{l}^{\mathrm {d}} \in \mathcal{CN}(0,\sigma^2 \mathbf{I}_M)$ is the additive white Gaussian noise. For signal detection in the first stage, the $l^\text{th}$ AP utilizes the maximum ratio combining technique in which the received signal, $\mathbf {r}_l^{\mathrm {d}}$ is multiplied by the Hermitian transpose of the channel estimate,  $\hat { \mathbf {h}}_{lu}^\mathrm{H}$ and transmits $\hat { \mathbf h}_{lu}^\mathrm{H} \mathbf {r}_l^{\mathrm {d}} $ to the CPU through the backhaul link for all users. Then, for each user, $u$, the received products are combined as
 \begin{align} \label{Recieved vector}
    r_u &= \sum_{l \in \mathcal{A}_u} \hat { \mathbf {h}}_{lu}^\mathrm{H} \mathbf {r}_l^{\mathrm {d}} \nonumber \\  &= \sum_{l \in \mathcal{A}_u} \Big( \hat{\mathbf{h}}_{lu}^\mathrm{H} {\mathbf{h}}_{lu} s_u + \sum_{i=1,i \neq u}^U \hat{\mathbf{h}}_{lu}^\mathrm{H} {\mathbf{h}}_{li} s_i + \hat{\mathbf{h}}_{lu}^\mathrm{H} \mathbf{n}_l^\mathrm{d} \Big).
 \end{align}
Since the CPU does not have the knowledge of the channel
estimates, we apply the use-and-then-forget technique \cite{marzetta2016fundamentals} to obtain the achievable SE as follows
\begin{equation} \label{Original Rate}
\text{SE}_{u} = (1 - \frac{\tau}{\tau _{\text {c}}}) \log _{2}\left ({1 +  \mathrm{SINR}_{u}}\right),
\end{equation}
where $\mathrm{SINR}_u$ is given at the top of the this page \cite{bjornson2017massive} and
\begin{align}
\mathbb{E}\left\{\hat{\mathbf{h}}_{lu}^{\mathrm{H}}  \mathbf{h}_{lu}\right\} & =p_u^\mathrm{p} \tau \sum_{l \in \mathcal{A}_u } \operatorname{tr}\left( \mathbf{B}_{lu} \boldsymbol{\Psi}_{l_{t_u}}^{-1} \mathbf{B}_{lu}\right). \\
\mathbb{E}\left\{\left\| \hat{\mathbf{h}}_{lu}\right\|^2\right\} & =p_u^\mathrm{p} \tau \sum_{l \in \mathcal{A}_u } \operatorname{tr}\left( \mathbf{B}_{lu} \boldsymbol{\Psi}_{l_{t_u}}^{-1} \mathbf{B}_{lu}\right).
\end{align}
and
\begin{align}
\mathbb{E} & \left\{\left|\hat{\mathbf{h}}_{lu}^{\mathrm{H}}  \mathbf{h}_{li}\right|^2\right\} =  p_u^\mathrm{p} \tau \sum_{l \in \mathcal{A}_u } \operatorname{tr}\left( \mathbf{B}_{l i} \mathbf{B}_{lu} \boldsymbol{\Psi}_{l_{t_u}}^{-1} \mathbf{B}_{lu}\right) \nonumber \\
& \quad+ \begin{cases}p_u^\mathrm{p} p_i^\mathrm{p} \tau^2\left|\sum_{l \in \mathcal{A}_u } \operatorname{tr}\left( \mathbf{B}_{l i} \boldsymbol{\Psi}_{l_{t_u}}^{-1} \mathbf{B}_{lu}\right)\right|^2 & i \in \mathcal{P}_u \\
0 & i \notin \mathcal{P}_u\end{cases}
\end{align}
\begin{figure*}[t]
\small
% \noindent\rule{\textwidth}{.5pt}%\vskip3pt
\begin{equation} \label{fin SINR}
\begin{aligned}
\operatorname{SINR}_u &=\frac{p_u^\mathrm{d}\left|\sum_{ l \in \mathcal{A}_u } \mathbb{E}\left\{\hat{\mathbf{h}}_{lu}^{\mathrm{H}} \mathbf{h}_{lu}\right\}\right|^2}{\sum_{i=1}^U p_i^\mathrm{d} \mathbb{E}\left\{\left|\sum_{ l \in \mathcal{A}_u } \hat{\mathbf{h}}_{lu}^{\mathrm{H}} \mathbf{h}_{l i}\right|^2\right\}-p_u^\mathrm{d}\left|\sum_{ l \in \mathcal{A}_u } \mathbb{E}\left\{\hat{\mathbf{h}}_{lu}^{\mathrm{H}} \mathbf{h}_{lu}\right\}\right|^2+\sigma^2 \sum_{ l \in \mathcal{A}_u } \mathbb{E}\left\{\left\|\hat{\mathbf{h}}_{lu}\right\|^2\right\}}  \\[6pt] &= \frac{ p_u^\mathrm{d} p_u^\mathrm{p} \tau 
\sum_{l \in \mathcal{A}_u } \operatorname{tr}( \mathbf{B}_{lu} \boldsymbol{\Psi}_{l_{t_u}}^{-1} \mathbf{B}_{lu}) }
{ \displaystyle 
\sum_{i=1}^U \sum_{l \in \mathcal{A}_u} 
   \frac{ p_i^\mathrm{d}\operatorname{tr}( \mathbf{B}_{l i} \mathbf{B}_{lu} \boldsymbol{\Psi}_{l_{t_u}}^{-1} \mathbf{B}_{lu})}
        {\operatorname{tr}( \mathbf{B}_{lu} \boldsymbol{\Psi}_{l_{t_u}}^{-1} \mathbf{B}_{lu})} 
+ \sum_{\substack{i \in \mathcal{P}_u \\ i \neq u}} \sum_{l \in \mathcal{A}_u} 
   \frac{p_i^\mathrm{p} p_u^\mathrm{d} \tau | \operatorname{tr}( \mathbf{B}_{l i} \boldsymbol{\Psi}_{l_{t_u}}^{-1} \mathbf{B}_{lu}) |^2}
        {\operatorname{tr}( \mathbf{B}_{lu} \boldsymbol{\Psi}_{l_{t_u}}^{-1} \mathbf{B}_{lu})} 
+ \sigma^2 }
\end{aligned}
\end{equation}\noindent\rule{\textwidth}{.9pt}%\vskip3pt
\end{figure*}
 
\section{Proposed AP selection and Pilot power allocation }
In this section, we present a novel, efficient, and scalable dynamic AP clustering method based on channel correlation, designed to ensure that low–mutually correlated APs serve users with the strongest channel conditions. In addition, we propose a pilot power allocation strategy to enhance each user’s SINR during the training phase. The proposed approaches address the weighted sum-rate maximization (WSRM) problem between APs and users.

 \subsection{Access Points Clustering} \label{AP clustering}
AP selection based on the channel gain is an efficient method in that it reduces complexity, backhaul load, and interference by choosing the APs with the strongest large-scale fading for each user. However, this approach also comes with certain drawbacks. One key limitation is determining the optimal number of APs to select for each user; choosing too few APs may degrade performance due to insufficient diversity or coverage, while selecting too many can negate the benefits of reduced complexity. Moreover, the selection based purely on channel gain does not consider AP correlation or spatial distribution, which may lead to suboptimal AP clusters and potential interference. On the other hand, to make a CFmMIMO system scalable, the system must
adopt an association scheme that limits the maximum number of UEs that can associate with one AP.

To address these challenges, we propose a dynamic clustering approach that groups APs based on the correlation of their channel responses for each user. This strategy ensures that the APs serving a given user exhibit similar channel characteristics, enabling optimization of the user’s signal reception. Furthermore, we introduce an adaptive AP–user association scheme designed to enhance scalability as the network grows. The overall procedure is organized into four key stages:\\
\textit{Stage 1: Channel Correlation \& AP Clustering }

In the proposed clustering formation, we aim to group a subset of the low-mutually correlated APs that provide similar channel conditions to the $u^\text{th}$ user dynamically. In this regard, the collective channel from all users to $l^\text{th}$ AP is defined as $\hat{\mathbf{h}}_l = [\hat{\mathbf{h}}_{l1}^\mathrm{T}, \hat{\mathbf{h}}_{l2}^\mathrm{T}, \ldots, \hat{\mathbf{h}}_{lU}^\mathrm{T}]^\mathrm{T} \in \mathbb{C}^{MU}$. Since we want to find the APs with the lowest correlation and similar channel characteristics that serve the $u^\text{th}$ user, the correlation coefficient between the channels of two APs, such as $l$ and $k$, can be expressed as:
\begin{align}
    \rho_{lk} = \frac{\hat{\mathbf{h}}_l^\mathrm{H}\hat{\mathbf{h}}_k}{ \| \hat{\mathbf{h}}_l \| \| \hat{\mathbf{h}}_k \|}, \quad \forall l,k \in \{ 1,2, \cdots, L\}
\end{align}
This evaluates the correlation between the channel vectors of different APs, ensuring that APs with low mutual correlation are grouped into the same cluster.  To this end, we utilize an algorithm that recursively merges or splits APs into nested clusters based on a measure of similarity, ultimately forming a tree-like structure.\\
We start by defining a distance matrix $\mathbf{D} \in \mathbb{R}^{L \times L}$ and its entries given by $D_{lk} = 1 - \rho_{lk}$. Note that the distance matrix does not represent the physical distances between pairs of APs. Instead, it quantifies the difference in their correlation values, providing a measure of similarity between the APs based on their channel characteristics. Each entry in the matrix corresponds to the computed "distance" in terms of correlation rather than geographic separation. After that, we utilize tree-like clustering to group APs based on their channel correlation. The clustering is based on the correlation coefficients $\rho_{lk}$ and the linkage function, which defines how to measure the distance between sets of data points. 
In this clustering, the process starts with each AP as an individual cluster as $\mathcal{C}^{(1)} = \{ \{AP_1\}, \{AP_2\}, \cdots,\{AP_L\} \}$, which is the initial clustering. Then, the inter-cluster distance $D(C_i,C_j)$ (i.e. $C_i= \{AP_i\}, \ C_j= \{AP_j\}$, \ $i\neq j$, $\forall i,j \in L $) between any two clusters using the linkage criterion is computed as
\begin{align}
    D(C_i,C_j) = \dfrac{1}{|C_i| |C_j|} \sum_{l \in C_i} \sum_{k \in C_j} D(l,k),
\end{align}
where $C_i$ and $C_j$ are the \textit{i}-th and \textit{j}-th current clusters ($j \neq i$). Then, the pair of clusters $(C_i, C_j)$ with minimum $D(C_i, C_j)$ is computed as $(i^*,j^*) = \text{arg} \min_{i \neq j} D(C_i, C_j)$. 
After that, the closest clusters are iteratively merged based on some similarity (or distance) measure as
\begin{align}
    \mathcal{C}^{(n+1)} = \Big( \mathcal{C}^{(n)} \backslash \{ C_{i^*}, C_{j^*}  \}  \Big) \cup \{ C_{i^*}, C_{j^*}  \},
\end{align}
where $n$ is the internal iteration of the linkage function. The clustering process continues iteratively until a predefined stopping condition is satisfied. Specifically, the merging of clusters is terminated when the inter-cluster distance exceeds a specified threshold \( \kappa \). Mathematically, let \( D(C_i, C_j) \) denote the distance between clusters \( C_i \) and \( C_j \), according to the chosen linkage criterion. The process halts when $\min_{i \ne j} D(C_i, C_j) > \kappa$. At this point, the current partition \( \mathcal{C} = \{ \mathcal{C}_1, \mathcal{C}_2, \dots, \mathcal{C}_S \} \) defines the final set of clusters, where $S$ is the total number of clusters.\\
\textit{Stage 2: AP-user Association}

Once the clusters are formed, AP selection proceeds as follows. For each user, the large-scale fading coefficients from all APs are sorted in descending order to identify the AP with the strongest channel gain. The algorithm then scans through the set of clusters to find the one that contains the strongest AP, $\mathcal{I}_u$. Once identified, the entire cluster containing the strongest AP is selected to serve the user. This ensures that the selected APs are not only strong individually but also form a spatially low-coherent group with respect to the user. \\
\textbf{Note:} If the strongest AP appears in more than one cluster (e.g., due to overlapping clusters from multiple users), a cluster resolution strategy is applied. In this case, we prioritize the cluster that maximizes the total sum of large-scale fading coefficients for the user. 
This helps avoid ambiguous AP assignments and ensures fair and interference-aware clustering. This approach provides a balanced trade-off: it retains the strongest APs for each user while forming clusters that promote coherent signal transmission and efficient interference management.

\textit{Stage 3: Scalability Constraint}

At this stage, the algorithm incorporates a load-balancing mechanism to prevent any AP from being overloaded with user associations. Each AP is allowed to serve at most $\tau$ users simultaneously. 
Before assigning a candidate cluster $\mathcal{C}_i$ to a user $u$, the algorithm checks the current load $N_l$ of every AP $l$ within that cluster. The assignment is accepted only if all APs in $\mathcal{C}_i$ satisfy the condition $N_l < \tau$. Once the assignment is confirmed, the cluster $\mathcal{C}_i$ is added to the user’s serving set $\mathcal{A}_u$, and the counters $N_l$ for the corresponding APs are incremented. 
This stage guarantees that the association process remains scalable with respect to network size and user density while keeping pilot contamination and interference under control.

\textit{Stage 4: Overflow Handling}

When the candidate cluster $\mathcal{C}_i$ for user $u$ contains one or more APs that have already reached their service limit $\tau$, the algorithm activates an overflow management mechanism to maintain balanced resource allocation. Two complementary strategies are employed: 
\textit{i. Partial Assignment:} The algorithm first filters out the saturated APs from $\mathcal{C}_i$ and retains only the subset of APs with available capacity, and this subset serves the user. 
\textit{ii. Cluster Reassignment:} If all APs in $\mathcal{C}_i$ are saturated and no valid subset remains, the algorithm searches for an alternative cluster $\mathcal{C}_{i^\text{alt}}$ with sufficient remaining capacity. The selection criterion is to maximize the aggregate channel gain $\sum_{l\in\mathcal{C}_{i^\text{alt}}}\beta_{lu}$ while respecting the constraint $N_l < \tau$. The user is then associated with this alternative cluster. 
This stage ensures fair distribution of network resources, prevents excessive load concentration on a few APs, and maintains service quality even in dense deployment scenarios.
The full procedure is outlined in Algorithm~\ref{alg_scalable}.
\begin{algorithm}[!]
\small
\setstretch{1.1} % single-spaced inside 
\caption{Proposed scalable method}
\label{alg_scalable}
\SetAlgoLined
\textbf{Input:} $\hat{\mathbf{h}}_{lu}$, $u=1, \cdots,U$, $\kappa$, $\tau$, $\mathcal{C}$: clustered APs, $\mathcal{A}_u$: set of APs serving user $u$ \;
\textbf{Initialization:} $\mathcal{A}_1 = \cdots = \mathcal{A}_U = \emptyset$; $N_l = 0, \forall l$ \\
\For{\text{channel realization}}{
  Define $\hat{\mathbf{h}}_l = [\hat{\mathbf{h}}_{l1}^\mathrm{T}, \ldots, \hat{\mathbf{h}}_{lU}^\mathrm{T}]^\mathrm{T} \in \mathbb{C}^{MU}$\;
  $\rho_{lk} = \frac{\hat{\mathbf{h}}_l\hat{\mathbf{h}}^\mathrm{H}_k}{ \| \hat{\mathbf{h}}_l \| \| \hat{\mathbf{h}}_k \|}$, $\mathbf{D} = 1 - \rho_{lk}$ \\
  Initialize $\mathcal{C}^{(1)} = \{ \{AP_1\}, \cdots,\{AP_L\} \}$\;
  \While{$\min_{i \ne j} D(C_i, C_j) > \kappa$}{
     $D(C_i,C_j) = \dfrac{1}{|C_i||C_j|} \sum\limits_{l \in C_i}\sum\limits_{k \in C_j} D(l,k)$\;
     $(i^*,j^*) = \arg\min_{i \neq j} D(C_i, C_j)$\;
     $\mathcal{C}^{(n+1)} = (\mathcal{C}^{(n)} \backslash \{C_{i^*},C_{j^*}\}) \cup \{C_{i^*} \cup C_{j^*}\}$\;
  }
  $\mathcal{C} = \{\mathcal{C}_1, \dots, \mathcal{C}_s, \cdots, \mathcal{C}_S\}$\;

  \For{$u = 1:U$}{
    $\text{APs}_u \leftarrow \text{argsort}_{l} (\hat{\mathbf{h}}_{lu})^{\downarrow}$\;
    $l^* \leftarrow \text{APs}_u(1)$  \;
    $\mathcal{I}_u \leftarrow \{ s \;|\; l^* \in \mathcal{C}_s \}$\;
    \eIf{$|\mathcal{I}_u| = 1$}{
        $i \leftarrow \mathcal{I}_u(1)$\;
    }{
        $i \leftarrow \arg\max_{s \in \mathcal{I}_u} \sum\limits_{l \in \mathcal{C}_s} \hat{\mathbf{h}}_{lu}$ \tcp*{\scriptsize highest total gain cluster}
    }

    \eIf{$N_l < \tau \;\; \forall l \in \mathcal{C}_i$}{
        $\mathcal{A}_u \leftarrow \mathcal{C}_i$\;
        \For{$l \in \mathcal{C}_i$}{ $N_l \leftarrow N_l + 1$ }
    }{ \tcp{\scriptsize partial assignment}
        $\mathcal{C}_{valid} \leftarrow \{ l \in \mathcal{C}_i \;|\; N_l < \tau \}$\  \;
        \eIf{$\mathcal{C}_{valid} \neq \emptyset$}{
            $\mathcal{A}_u \leftarrow$  $l \in \mathcal{C}_{valid}$\;
            \For{$l \in \mathcal{A}_u$}{ $N_l \leftarrow N_l + 1$ } 
        }{\tcp{\scriptsize reassignment} \For{$s = 1:S$}{
                $\text{gain}(s) = \sum\limits_{l \in \mathcal{C}_s, N_l < \tau} \hat{\mathbf{h}}_{lu}$\;
            } 
            $i^\text{alt} = \arg\max_s \text{gain}(s)$ \tcp{\scriptsize alternative} 
            $\mathcal{A}_u \leftarrow \{ l \in \mathcal{C}_{i^\text{alt}} \;|\; N_l < \tau \}$\;
            \For{$l \in \mathcal{A}_u$}{ $N_l \leftarrow N_l + 1$ }
        }
    }
  }
}
\textbf{Output:} $\mathcal{A}_u$ for all $u$
\end{algorithm}

\subsection{Pilot power allocation}
The CFmMIMO system operates in TDD mode, with channels estimated at each AP based on user pilot signals during the training phase. However, due to the non-orthogonality of pilot sequences, the channel estimates for a given user are corrupted by interference from pilots sent by other users, a phenomenon known as pilot contamination \cite{ngo2017cell}. This issue is particularly pronounced where the goal is to simultaneously serve a large number of users within the same time-frequency resources. As a result, mitigating pilot contamination is critical. To tackle pilot contamination, recent studies have introduced various algorithms aimed at mitigating its effects, especially in densely populated multi-user settings, to improve communication quality by reducing interference from pilot reuse \cite{gottsch2022subspace, osawa2023overloaded}.\\
However, this work assumes a scenario with substantial pilot contamination resulting from random pilot assignment. This assumption enables the evaluation of the proposed method’s performance under challenging, worst-case conditions. Thus, we propose a solution optimizing pilot power allocations to maximize each user's SINR during the training phase.

\subsubsection{Weighted sum-rate maximization problem}

Given the objective of providing a better quality of service to all users, we formulate this as a WSRM problem \cite{weeraddana2012weighted}. The system dynamically adjusts pilot powers to maximize the weighted sum of users' data rates subject to power constraints as follows
 \begin{align} \label{WSR}
P_1: \quad   \max _{\mathbf{p}_u^\mathrm{p}} \sum _{u=1}^U w_u \text{SE}_{u}(\mathbf{p}_u^\mathrm{p},\mathbf{p}_u^\mathrm{d})  \quad \text{s.t.}  \quad  \epsilon \leq  p_u^\mathrm{p}  \leq  P_{\text{max}} \quad \forall u ,
\end{align}
where $w_u$ is a weight associated with the $u^\text{th}$ user (representing user priority) and $P_{\text{max}}$ denotes the maximum transmit budget available for each user while $\epsilon > 0$ is small value. Also, we have assumed that the data power, $\mathbf{p}_u^\mathrm{d}$, is fixed and distributed between all users equally.
It is clear from the optimization problem in \eqref{WSR} that the SE for each user $u$ depends on their SINR, which is a function of the power allocation as follows
\begin{align} \label{P1}
P_1: \quad  & \max _{\mathbf{p}_u^\mathrm{p}} \sum _{u=1}^U w_u \log _{2}\left ({1 +  \mathrm{SINR}_{u}(\mathbf{p}_u^\mathrm{p}, \mathbf{p}_u^\mathrm{d})}\right),\nonumber  \\ & \quad \text{s.t.}  \quad  \epsilon \leq  p_u^\mathrm{p}  \leq  P_{\text{max}} \quad \forall u ,
\end{align}
Although the problem $P_1$ is an NP-hard problem \cite{gruzdeva2018solving}, it can be solved using a quadratic transform for multiple-ratio fractional programming \cite{shen2018fractional}.
In this problem, it is seen that the power control problem is not explicitly in a ratio form, but the key components of the objective function, namely, the SINR terms, exhibit a fractional structure. Since each SINR expression appears within a logarithmic function, which is concave and non-decreasing, we can solve it using the following corollary.\\
\textit{Corollary 1:} Given a sequence of non-decreasing functions $f_m(\cdot)$ and a sequence of ratios $A_m / B_m$ for $m=1, \ldots, M$, the sum-of-functions-of-ratio problem
\begin{align} \label{corollary1}
% \begin{array}{ll}
\underset{\mathbf{x}}{\operatorname{max}} & \sum_{m=1}^M f_m\left(\frac{A_m(\mathbf{x})}{B_m(\mathbf{x})}\right)  \quad 
\text { s.t. } \quad \mathbf{x} \in \mathcal{X}
% \end{array}
\end{align}
is equivalent to
\begin{subequations}\label{11}
\begin{align} 
\underset{\mathbf{x}, \mathbf{y}}{\operatorname{max}} \quad & \sum_{m=1}^M f_m\left(2 y_m \sqrt{A_m(\mathbf{x})}-y_m^2 B_m(\mathbf{x})\right) \label{1} \\
\text{s.t.} \quad & \mathbf{x} \in \mathcal{X}, \ \  y_m \in \mathbb{R}, \quad m = 1, \ldots, M \label{2}
\end{align}
\end{subequations}
To verify Corollary 1, problem \eqref{corollary1}  is reformulated as
\begin{align}
    \max_{\mathbf{x}, \mathbf{r}} \; \sum_{m=1}^M f_m(r_m) \quad \text{s.t. \quad  } \mathbf{x} \in \mathcal{X}, \; r_m = \frac{A_m(\mathbf{x})}{B_m(\mathbf{x})}
\end{align}
Applying the quadratic transform introduced in \cite{shen2018fractional}, each ratio term $r_m$ can be equivalently expressed as
\begin{align}
    r_m = \max_{y_m} \Big( 2 y_m \sqrt{A_m(\mathbf{x})} - y_m^2 B_m(\mathbf{x}) \Big),
\end{align}
where $y_m$ is the auxiliary available. Since $f_m(\cdot)$ is nondecreasing, the problem becomes
\begin{align} \label{maximize}
    \max_{\mathbf{x}} \sum_{m=1}^M f_m\Big( \max_{y_m} \big( 2 y_m \sqrt{A_m(\mathbf{x})} - y_m^2 B_m(\mathbf{x}) \big) \Big).
\end{align}
Finally, by combining the outer maximization over $\mathbf{x}$ and the inner maximization over $\mathbf{y}$, we obtain the equivalent formulation given in \eqref{1}.\\
\indent The optimization problem in \eqref{11} can be solved using an iterative approach by applying the quadratic transform, which alternates between optimizing the primal variable $\mathbf{x}$ and updating $y_m$ in closed form as follows:\\
When we hold $\mathbf{x}$ fixed, the optimal $y_m$ that maximizes the objective function can be obtained by taking the first-order derivative of \eqref{maximize} and setting it to zero
\begin{align} \label{y star}
  y_m^{\star}=\frac{\sqrt{A_m(\mathbf{x})}}{B_m(\mathbf{x})}, \forall m=1, \cdots, M.  
\end{align}
When $y_m$ is fixed, due to the concavity of each $A_m(\mathbf{x})$, the convexity of each $B_m(\mathbf{x})$, and the fact that the square-root function is concave and increasing, the quadratic transform
\begin{align}
  g \left(\mathbf{x}, y_m\right)=2 y_m \sqrt{A_m(\mathbf{x})} - y_m^2 B_m(\mathbf{x}),
\end{align}
is concave in $\mathbf{x}$. Further, if $f_m(\cdot)$ is assumed to be concave and nondecreasing, then we also have that $f_m\left(g\left(\mathbf{x}, y_m\right)\right)$ is concave in $\mathbf{x}$. Therefore, the quadratic transformed problem \eqref{1} is a concave maximization problem over $\mathbf{x}$. The optimal $\mathbf{x}$ can thus be efficiently obtained through numerical convex optimization. The entire approach is summarized in Algorithm~\ref{alg_2} \cite{shen2018fractional}.\\
\begin{algorithm}[t]
\small
\caption{Iterative Approach for Concave-Convex fractional problem in \eqref{corollary1}}
\label{alg_2}
\SetAlgoNoLine
\textbf{Step 0:} Initialize $\mathbf{x}$ to a feasible value \;
 \textbf{Step 1} Reformulate the problem by the quadratic transform,
i.e., replace each $A_m / B_m$ with $2y_m \sqrt{A_m} - y_m^2 B_m$ \;
   \Repeat{convergence}{ \textbf{Step 2:} Update $\mathbf{y}$ by \eqref{y star}
   \\ \textbf{Step 3:} Update $\mathbf{x}$ by solving the reformulated convex
optimization problem \eqref{1},
over $\mathbf{x}$ for fixed $\mathbf{y}$  }   
\end{algorithm}
It can be shown by the below theorem that Algorithm~\ref{alg_2} will achieve a stationary point of concave-convex fractional programming problems.\\
\textit{Theorem 1:} For the concave-convex sum-of-functions-of ratio problem \eqref{corollary1}, i.e., every $A_m (\mathbf{x})$ is concave and every $B_m (\mathbf
x)$ is convex, and $\mathcal{X}$ is a convex set in standard form, and assuming further that $f_m$ is non-decreasing and concave, then Algorithm~\ref{alg_2} consists of a sequence of convex optimization problems that converge to a stationary point of \eqref{corollary1} with non-decreasing sum-of-functions-of-ratio value after every iteration.\\
\textit{Proof:} The algorithm is essentially a block coordinate ascent algorithm for the reformulated problem \eqref{11}, which is a convex optimization problem due to the concave-convex form of \eqref{corollary1}, so it converges to a stationary point ($\mathbf{x}^*, \mathbf{y}^*$) of \eqref{11}. Due to the equivalence in the solution and the equivalence in the objective value, the first-order condition on $\mathbf{x}$ for \eqref{11} under the optimal $\mathbf{y}$ is the same as for the original problem \eqref{corollary1}, hence the algorithm also converges to a stationary point of \eqref{corollary1}. Also, the sum-of-ratio function is nondecreasing after each $\mathbf{y}$ update.\\
Using this theorem and according to \eqref{11}, the problem in $P_1$ can be written as
\begin{subequations}
\begin{align} 
P_2: \quad   \max_{\{p_u^\mathrm{p}, y_u\}} &\sum _{u=1}^U w_u \Big(2 y_m \sqrt{A_u(p_u^\mathrm{p})}-y_m^2 B_u(p^\mathrm{p}_{-u})\Big)   \\   \quad  &\epsilon \leq  p_u^\mathrm{p}  \leq  P_{\text{max}}, \quad  \quad u = 1, \ldots, U\\
& y_i \in \mathbb{R}, 
\end{align}
\end{subequations}
From \eqref{fin SINR}, the terms $A_u(p_u^\mathrm{p})$ and $B_u(p_{-u}^\mathrm{p})$ correspond to the numerator and denominator, respectively.
Problem 2 ($P_2$) is solved through an iterative procedure, as outlined in Algorithm~\ref{alg:P3}.

\begin{algorithm}[!]
\small
\caption{Power Control Approach}
\label{alg:P3}
\SetAlgoLined
\KwIn{Initial pilot powers $\{p_u^\mathrm{p}(0)\} \in [\epsilon,P_{\max}]$, weights $\{w_u\}$, tolerance $\delta$, max iterations $N_{\max}$}
\KwOut{Optimized pilot powers $\{p_u^\mathrm{p*}\}$}

\textbf{Initialize:} $p_u^\mathrm{p}(0) \leftarrow P_{\max}/2, \forall u$, iteration $n \leftarrow 0$ \;

\Repeat{ $\frac{\|\mathbf{p}^\mathrm{p}(n)-\mathbf{p}^\mathrm{p}(n-1)\|}{\|\mathbf{p}^\mathrm{p}(n-1)\|} < \delta$   }{

    \tcp{\scriptsize Step 1: Update auxiliary variables $y_u$ }
    \For{$u=1:U$}{
        $y_u \leftarrow \frac{\sqrt{A_u(p_u^\mathrm{p})}}{B_u(p_{-u}^\mathrm{p})}$ \;
    } 
    \tcp{\scriptsize Step 2: Update pilot powers $\{p_u^\mathrm{p}\}$ }
    Solve the following problem via CVX \cite{grant2008cvx}:
    \begin{align*}
    \max_{\{p_u^\mathrm{p}\}} \; & \sum_{u=1}^U w_u \Big( 2 y_u \sqrt{A_u(p_u^\mathrm{p})} - y_u^2 B_u(p_{-u}^\mathrm{p}) \Big) \qquad \nonumber \\ &
    \text{s.t.} \quad \epsilon \leq p_u^\mathrm{p} \leq P_{\max}, \quad \forall u \, .
    \end{align*}

    \textbf{Update:} $\mathbf{p}^\mathrm{p}(n+1) \leftarrow \{p_u^\mathrm{p}\}$ \;
    $n \leftarrow n+1$ \;
}
\Return{$\{p_u^\mathrm{p*}\} \leftarrow \mathbf{p}^\mathrm{p}(n)$}
\end{algorithm}

\begin{figure}[!]
	\centering 
	\includegraphics[width=3.5in,height=2.1in]{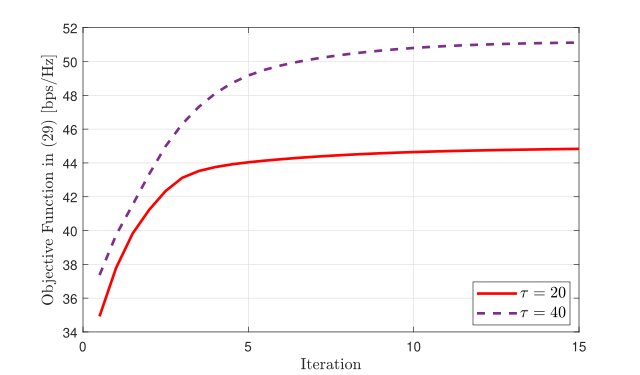}
	\caption{Convergence of the iterative algorithm for  $L=100, M=1, U=40$.} \label{Iteration figure}
\end{figure} 
Fig.~\ref{Iteration figure} depicts the convergence performance of the proposed power allocation scheme for different pilot sequences with $U = 40$ users and $L=100$ APs. 

\subsection{Data Power Control}

One important aspect of CFmMIMO systems is their capability to provide consistently uniform service across users. We focus on the optimization problem of max-min fairness, which entails using the pilot power coefficients calculated in Algorithm~\ref{alg_scalable} ($\hat{\mathbf{p}}^{\mathrm{p}}_u$) and optimizing the data power control coefficient to maximize the minimum user rates. We denote the data power coefficient vector as $\mathbf{p}_u^\mathrm{d} = [p_1^\mathrm{d}, p_2^\mathrm{d}, \cdots, p_U^\mathrm{d}]$, then rewrite the uplink SE of $u^\text{th}$ user  as
\begin{equation} \label{Original Rate}
\text{SE}_{u} (\hat{\mathbf{p}}_u^\mathrm{p},\mathbf{p}_u^\mathrm{d}) = (1 - \frac{\tau}{\tau _{\text {c}}} )\log _{2}\left ({1 +  \mathrm{SINR}_{u} (\hat{\mathbf{p}}_u^\mathrm{p},\mathbf{p}_u^\mathrm{d})}\right),
\end{equation}
and the optimization problem is formulated as:
\begin{align} \label{original data}
P_3: \quad &\max _{\lbrace p_u^\mathrm{d}\rbrace } \, \min _{u=1, \ldots, U} \text{SE}_{u} (\hat{\mathbf{p}}_u^\mathrm{p},\mathbf{p}_u^\mathrm{d})  \quad \text{s.t.} \quad 
\begin{aligned}[t]
& 0 \leq p_u^\mathrm{d} \leq 1, \quad \forall u \in U 
\end{aligned}
\end{align}
where $\text{SE}_{u} (\hat{\mathbf{p}}_u^\mathrm{p},\mathbf{p}_u^\mathrm{d})$ is defined in \eqref{Original Rate}. The optimal data power, ${p}_u^\mathrm{d}$ can be determined by solving the following max-min problem:
\begin{align} 
P_4: \quad &  \max _{\lbrace p_u^\mathrm{d}\rbrace } \, \min _{u=1, \ldots, U} \mathrm{SINR}_{u} (\hat{\mathbf{p}}_u^\mathrm{p},\mathbf{p}_u^\mathrm{d}) \quad  \text{s.t \quad }   0 \leq p_u^\mathrm{d}  \leq 1,   \quad \forall u \in U \label{zeroone}, 
\end{align}
Without loss of generality, problem $P_3$ can be rewritten by introducing a new slack variable as
\begin{subequations}\label{GP}
\begin{align}
P_4: \quad   \max _{{\lbrace p_k^\mathrm{d}\rbrace },t} \, t  \nonumber \\
 \text{s.t \quad } &t \leq \mathrm{SINR}_{u} (\hat{\mathbf{p}}_u^\mathrm{p},\mathbf{p}_u^\mathrm{d}), \label{SINR}\\
  &0 \leq p_u^\mathrm{d}  \leq 1,   \quad \forall u \in U \label{zeroone}, 
\end{align}
\end{subequations}
To tackle this optimization problem and explicitly expand the $\mathrm{SINR}_u$ expression in \eqref{fin SINR}, we focus on the special case of spatially uncorrelated fading, i.e., $\mathbf{B}_{lu}=\beta_{lu} \mathbf{I}_M$. Under this assumption, \eqref{fin SINR} simplifies to
\begin{align}
\dfrac{M \tau p_u^\mathrm{d} p_u^\mathrm{p}  \sum_{l \in \mathcal{A}_u}\beta_{lu}^2 {\psi}_{l}^{-1} }{ \sum_{l \in \mathcal{A}_u} \big( M \tau \sum _{\substack{i \in \mathcal{P}_u\\ i\ne u}} p_i^\mathrm{d} p_i^\mathrm{p} \beta_{li}^2 {\psi}_{l}^{-1} +  \sum_{i=1}^U p_i^\mathrm{d}  \beta_{li} \big) + \sigma^2}
\end{align}
where $\psi_{l} = \tau \sum_{i \in \mathcal{P}_u} p_i^\mathrm{p} \beta_{li} +\sigma^2 $.   \\[6pt]
\textbf{Proposition 2:} The problem \( P_4 \) can be reformulated as a geometric program (GP) \cite{chiang2007power}.\\
\textit{Proof:} Based on the standard form of a geometric progression  in \cite{boyd2004convex} we can reformulate the SINR constraint in \eqref{SINR} as
\begin{align}
    \dfrac{\sum_{l \in \mathcal{A}_u} \big( M \tau \sum _{\substack{i \in \mathcal{P}_u\\ i\ne u}} p_i^\mathrm{d} p_i^\mathrm{p} \beta_{li}^2 \psi_{l}^{-1} +  \sum_{i=1}^U p_i^\mathrm{d}  \beta_{li}  + \sigma^2\big) }{M \tau p_u^\mathrm{d} p_u^\mathrm{p}  \sum_{l \in \mathcal{A}_u}\beta_{lu}^2 \psi_{l}^{-1} } < \dfrac{1}{t}
 \end{align}
With a straightforward transformation, the SINR constraint in \eqref{SINR} can be expressed as the following inequality:
\begin{align}\label{Transform}
    \frac{1}{p_u^\mathrm{d}} \Big(\sum_{\substack{ i \in \mathcal{P}_u \\i \ne u}}^{U} e_{ui}p_i^\mathrm{d} + \sum_{i=1}^U f_{ui} p_i^\mathrm{d} + b_u \Big) < \dfrac{1}{t}
\end{align}
where
\begin{align}
    e_{ui} &= \frac{\sum_{l \in \mathcal{A}_u} p_i^\mathrm{p} \beta_{li}^2 }{ p_u^\mathrm{p} \sum_{l \in \mathcal{A}_u} \beta_{lu}^2 } , \quad 
    f_{ui} = \dfrac{\sum_{l \in \mathcal{A}_u} \beta_{li}}{M \tau p_u^\mathrm{p} \sum_{l \in \mathcal{A}_u} \beta_{lu}^2 \psi_{l}^{-1}} \quad \nonumber  \\ &
    b_u = \frac{\sigma^2}{M \tau p_u^\mathrm{p} \sum_{l \in \mathcal{A}_u} \beta_{lu}^2 \psi_{l}^{-1}}
\end{align}
The transformation in \eqref{Transform} reveals that the left-hand side is a posynomial function. Consequently, problem \( P_4 \) meets the criteria for a standard GP problem, thereby validating the proof of Proposition 2. As stated in Proposition 2, the objective function and constraints of \( P_4 \) consist of monomial and posynomial functions with respect to the power allocations \( p_u^\mathrm{d} \). This confirms that \( P_4 \) is a standard GP, which can be efficiently solved using convex optimization tools.

\subsection{Complexity Analysis of the Proposed Method}

The computational complexity of the scalable AP-user association algorithm is mainly determined by three components. First, computing the AP--AP correlation matrix requires evaluating the correlation coefficient for all $L(L-1)/2$ AP pairs, each involving an inner product over $U$ users, leading to a complexity of $\mathcal{O}(L^2 U)$. Second, the hierarchical clustering stage, implemented via agglomerative merging, requires identifying and merging the closest cluster pair at each iteration, which in the naive implementation leads to $\mathcal{O}(L^2)$ operations in the worst case. Third, ranking the APs for each user based on channel gains involves sorting $L$ values per user, yielding $\mathcal{O}(U L \log L)$ complexity. Final overall complexity is $\mathcal{O}(L^2 + L^2 U + U L \log L)$, which simplifies to $\mathcal{O}(L^2)$. \\
\indent The computational complexity of the proposed power control algorithm (Algorithm~\ref{alg:P3}) is primarily determined by the iterative nature of the method and the convex optimization solved at each iteration. In each iteration, the auxiliary variables \( y_u \) for all \( U \) users are updated with a complexity on the order of \( \mathcal{O}(U^2) \), assuming the computation of functions \( A_u(\cdot) \) and \( B_u(\cdot) \) involves summations over all users. The dominant computational cost arises from solving the convex optimization problem via CVX, which involves \( U \) variables and has a worst-case complexity approximately \( \mathcal{O}(U^3) \) when using interior-point methods. Considering that the algorithm runs for at most \( N_{\max} \) iterations until convergence, the overall complexity can be expressed as \( \mathcal{O}(N_{\max} U^3) \). This cubic complexity with respect to the number of users reflects the computational demand of the pilot power optimization step in CFmMIMO systems, while the optimization solver for the data power problem $P_3$ in \eqref{original data} incurs complexity of $\mathcal{O}(U^{3.5})$.
\add{The computational complexity of the compared approaches is summarized in Table~\ref{Com Complexity}. 
\begin{table}[h]
\centering
\captionsetup{labelformat=empty}
\caption{TABLE I
Complexity Comparison of the Benchmarks.}
\begin{tabular}{|c|c|}
\hline Schemes & Computational complexity \\
\hline \hline DAPPA & $\mathcal{O}\big( U^{3.5}+ |\mathcal{A}_u|^2\big)$ \\
\hline GB \cite{liu2020graph} & $\mathcal{O}\left(U^3L^2M^2+ U^2LM\tau + U M \log _2 M\right)$ \\
\hline DCC \cite{bjornson2020scalable} & $\mathcal{O}\big( (M\tau + M^2) |\mathcal{A}_u|^2U  \big)$ \\
\hline NEEPC \cite{sarker2023pilot} &$\mathcal{O}\big( \mathcal{I}_\text{outer}(\max(  (27U^{3.5}), 9(M+5)U^2)\big)$
\\
\hline DLPC \cite{zhang2021deep} & $\mathcal{O}\left(L^3 +L^2  U M \log _2 L\right)$ 
\\
\hline SAT \cite{mai2018pilot} & $\mathcal{O}(\mathcal{I}_\text{outer} \cdot \mathcal{I}_\text{inner} \cdot U^4)$ \\
\hline
\end{tabular}
\label{Com Complexity}
\end{table}}\\
\add{where $\mathcal{I}_\text{outer}$ and $\mathcal{I}_\text{inner}$ are the outer and inner iterations of the methods in \cite{sarker2023pilot} and \cite{mai2018pilot}.}

\section{Numerical Results}
\subsection{Propagation Model}
\begin{figure}[!]
	\centering
	\includegraphics[width=3.6in,height=2.1in]{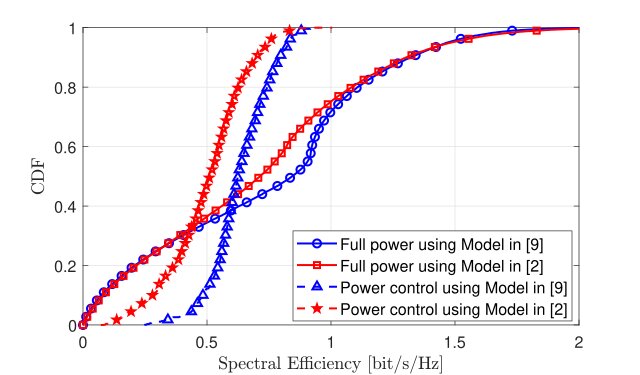}
	\caption{SE comparison of the proposed method in two propagation models in \cite{ngo2017cell} and \cite{bjornson2020scalable} for $L=100$, $M=1$, $U=40$, and $\tau=20$.}
 \label{Two Models}
\end{figure}

This section presents numerical results to assess the performance of the proposed approach under various CFmMIMO scenarios. While many recent studies on uplink maximum ratio combining have adopted the propagation model from \cite{ngo2017cell}, it is important to recognize that this model is derived from the COST-Hata framework in \cite{mogensen1999cost}, originally designed for macro-cell environments. Specifically, it assumes that APs are deployed at heights of at least 30 meters, and UEs are located at distances exceeding 1 $\text{km}$ from the APs. These assumptions significantly differ from the dense, micro-cell-like deployments expected in CFmMIMO systems. Notably, the authors of the COST-Hata model explicitly indicated that it “must not be used for micro-cells” \cite{bjornson2019making}. To ensure a fair and meaningful performance evaluation, we consider both propagation models from \cite{ngo2017cell} and \cite{bjornson2020scalable}, and select the most appropriate model based on comparative simulation results, as illustrated in Fig.~\ref{Two Models}.\\
\indent We consider simulation setups with $L$ APs with $M$ antennas, and $U$ single antenna users. Each coherence block contains $\tau_c = 200$ samples, and the total number of distributed antenna elements across the network is given by $LM$. Unless otherwise stated, $\tau = 20$ symbols are reserved for uplink pilot transmission, and each user uses its full transmit power $ 100 \ \mathrm{mW}$ for pilot signaling. We performed 1000-channel realizations. It is assumed that APs and users are independently and uniformly distributed within an area of $1 \times 1$ $\text{km}^2$, and a wrap-around technique is used to prevent boundary effects and simulate a network with an infinite area. The following three-slope propagation model was used in \cite{ngo2017cell}:
\begin{align}
\small
\begin{aligned}
 \beta_{lu} 
& = \begin{cases}-81.2, & d_{lu}<d_0  \\
-61.2-20 \log _{10}\left(\frac{d_{lu}}{1 }\right), & d_0  \leq d_{lu}<d_1  \\
-35.7-35 \log _{10}\left(\frac{d_{lu}}{1 }\right)+F_{lu}, & d_{lu} \geq d_1 
\end{cases}
\end{aligned}
\end{align}
where \(d_0 = 10 \mathrm{~m}\), \(d_1 = 50 \mathrm{~m}\) and $d_{lu}$ is the horizontal distance between the user $u$ and AP $l$ (i.e., ignoring the height difference). The shadowing term $F_{lu} \sim \mathcal{CN}\left(0,8^2\right)$ only appears when the distance is larger than 50 m and the terms are correlated as
\begin{align}\small
    \mathbb{E}\left\{F_{lu} F_{ji}\right\}=\frac{8^2}{2}\left(2^{-\delta_{u i} / 100 }+2^{-\varrho_{l j} / 100 }\right)
\end{align}
where $\delta_{u i}$ is the distance between UE $u$ and UE $i$ and $\varrho_{lj}$ is the distance between AP $l$ and AP $j$. The maximum user power is 100 mW, the bandwidth is 20 MHz, and the noise power is $92 \ \mathrm{dBm}$.
The large-scale fading coefficient (channel gain) in \cite{bjornson2020scalable} is modeled as:
\begin{align}
\small
   \beta_{lu}=-30.5-36.7 \log _{10} \big(\frac{d_{lu}}{1 }\big)+F_{lu}
\end{align}
where $d_{lu}$ is the three-dimensional distance between AP $l$ and the user $u$. The APs are deployed 10 m above the plane where the users are located, which acts as the minimum distance. This model matches the 3GPP Urban Microcell model in \cite{3gpp2017ts36814}. The shadow fading is $F_{lu} \sim \mathcal{N}\left(0,4^2\right)$ and the terms from an AP to different users are correlated as \cite{3gpp2017ts36814}
\begin{align}\small
  \mathbb{E}\left\{F_{lu} F_{ji}\right\}= \begin{cases}4^2 2^{-\delta_{u i} / 9 } & l=j \\ 0 & l \neq j\end{cases}  
\end{align}
As shown in Fig.~\ref{Two Models}, due to the limitations of the COST-Hata-based model in \cite{ngo2017cell} for micro-cell scenarios—and in accordance with the recommendation in \cite{bjornson2019making}—this work adopts the propagation model introduced in \cite{bjornson2020scalable}. This model is better suited for dense and distributed AP deployments in CFmMIMO systems and offers a more realistic assessment of system performance.
\subsection{Simulation Results}
\add{In this section, we evaluate the performance of the proposed power allocation and AP selection strategies detailed in Algorithms~\ref{alg_scalable} and~\ref{alg:P3}. The simulation results are organized into three subsections: the first examines the AP selection method using Algorithm~\ref{alg_scalable}; the second compares the power allocation strategy from Algorithm~\ref{alg:P3} against alternatives when all APs are active; and the third evaluates the combined performance of AP selection and power allocation under the proposed scheme.}
 
\subsubsection{Results based on the AP selection and the equal power allocation}

\begin{figure}[!]
	\centering
\includegraphics[width=3.6in,height=2.1in]{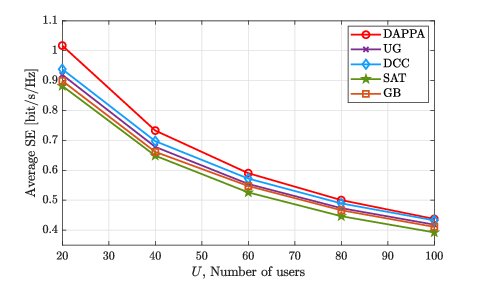}
	\caption{Impact of increasing user density on the average SE performance of different AP selection schemes with $\tau = 20$, $L = 100$ APs, $M = 1$ antenna, and varying numbers of users.}
 \label{Ave. ve. Users}
\end{figure}

In the first part of the simulation, we compare the proposed AP selection method with the scenario where only AP selection is considered, and the power is distributed equally among users. The proposed dynamic access point selection and pilot power allocation (DAPPA) is compared with well-known techniques such as dynamic cooperation clustering (DCC) in \cite{bjornson2020scalable}, user-group based pilot assignment (UG) in \cite{chen2020structured}, graph-based (GB) AP clustering in \cite{liu2020graph}, and successive approximation technique (SAT) in \cite{mai2018pilot}.
\begin{figure}[!]
	\centering
\includegraphics[width=3.6in,height=2.1in]{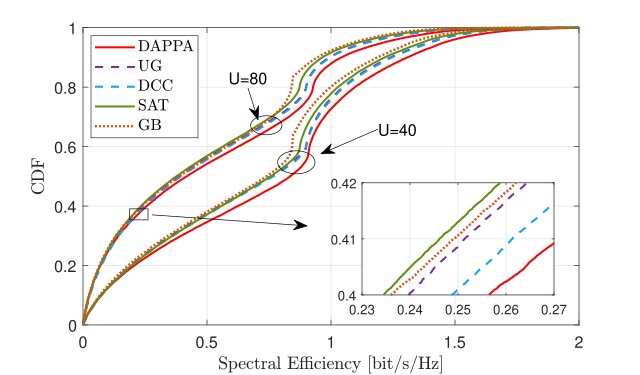}
	\caption{SE performance for different AP selection schemes with uniform power control with $\tau=20$, $L = 100$ APs, $M=1$ antennas, $U = 40$ and $U = 80$ users}
 \label{CDF vs Rate}
\end{figure} 
Fig.~\ref{CDF vs Rate} depicts the cumulative distribution function (CDF) of the SE for five AP selection schemes under uniform power allocation (It means that, the pilot signals are transmitted by all UEs with full power for all considered schemes), with $\tau = 20$, $L = 100$ APs, $M = 1$ antenna, and two different user counts, $U = 40$ and $U = 80$. For both user scenarios, the proposed method consistently indicates higher SE across most percentiles. In the $U = 40$ case, the proposed scheme offers. The proposed scheme delivers substantial SE improvements for nearly 80\% of the users, outperforming all other approaches. The inset zoom highlights that, at low SE values, the proposed scheme yields marginally higher performance, suggesting better service for far users. For $U = 80$, all schemes exhibit reduced SE due to increased interference and resource contention, yet the proposed method maintains superiority, particularly in the median SE region, where it surpasses the DCC method by a visible margin. UG and DCC remain competitive with each other, while GB slightly outperforms SAT in most ranges. Overall, the results show that the proposed approach given in Algorithm~\ref{alg_scalable} delivers a dynamic clustering approach that groups APs based on the correlation of their channel responses for each user. It demonstrates robust gains across the SE distribution for both moderate and dense user deployments.\\
Fig.~\ref{Ave. ve. Users} compare the average SE performance of different AP selection schemes for $\tau = 20$, $L = 100$ APs, and $M = 1$ antenna, as the number of users $U$ increases from 20 to 100, thereby capturing the impact of growing user congestion.  Across all user loads, the proposed DAPPA scheme achieves the highest SE, starting at approximately $1.05 \ \mathrm{bit/s/Hz}$ for $U=20$ and gradually decreasing to around $0.43 \ \mathrm{bit/s/Hz}$ at $U=100$. The UG and DCC schemes perform similarly, with DCC slightly outperforming UG at lower $U$ but converging as $U$ increases. SAT and GB consistently yield the lowest SE, with SAT generally performing slightly worse than GB for all $U$. The performance gap between the proposed scheme and the others is most pronounced for small $U$ (around $7\%$ improvement over the next best at $U=20$), and although this gap narrows with increasing $U$, the proposed method retains a noticeable advantage even in high-user scenarios. This trend indicates that while all methods experience performance degradation due to increased inter-user interference and limited pilot resources, the proposed scheme is more resilient, offering superior SE across the entire range of user densities.

\begin{figure}[!]
	\centering
\includegraphics[width=3.6in,height=2.1in]{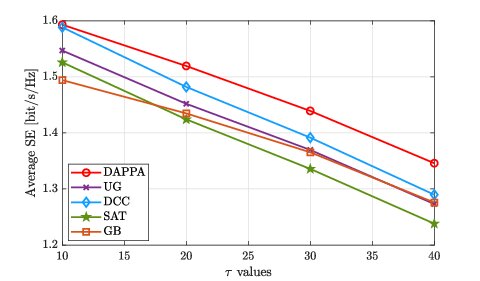}
	\caption{Average SE performance for different AP selection schemes and the number of pilot sequences with $L = 100$ APs, $M=4$ antennas, $U = 50$ users,
and $\tau_c = 200$}
 \label{Ave. ve. tau}
\end{figure} 

\begin{figure}[!]
	\centering
\includegraphics[width=3.6in,height=2.1in]{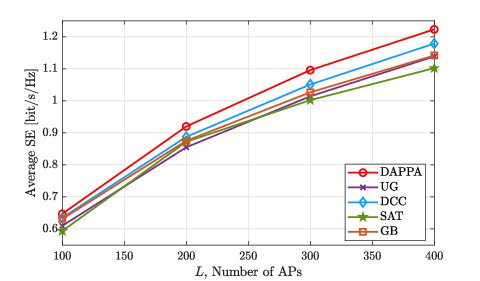}
	\caption{AP configuration's impact on average SE performance with $\tau=20$, $U = 50$ user, $M=1$ antennas, and different number of APs, $L$}
 \label{Ave. ve. APs}
\end{figure}

Fig.~\ref{Ave. ve. tau} illustrates the impact of the number of pilot sequences $\tau$ on the average SE performance of various AP selection techniques, considering $L = 100$ APs, $M = 4$ antennas, $U = 50$ users, and $\tau_c = 200$.
Across all $\tau$ values, the proposed DAPPA method consistently achieves the highest SE, starting at approximately $1.60\ \mathrm{bit/s/Hz}$ for $\tau = 10$ and gradually decreasing to about $1.36\ \mathrm{bit/s/Hz}$ at $\tau = 40$. The DCC and UG schemes follow closely, with DCC exhibiting slightly better performance than UG for smaller $\tau$ values, but converging at larger $\tau$. GB and SAT produce the lowest SE results, with GB marginally outperforming SAT in most scenarios. The performance gap between the proposed method and the others is most pronounced at $\tau = 10$ (approximately $4$--$6\%$ higher SE compared to DCC and UG) and narrows as $\tau$ increases, suggesting that the proposed method gains a greater advantage under shorter pilot sequences, where pilot contamination effects are more severe. All schemes exhibit a monotonic decrease in SE with increasing $\tau$, primarily due to the reduced duration available for data transmission within the coherence interval. The results indicate that the average SE reaches its maximum at $\tau = 10$ for all methods and then declines. This behavior can be explained by the fixed total power per coherence block: distributing this power over longer pilot sequences lowers the per-symbol pilot power, thereby degrading channel estimation accuracy. Furthermore, according to the SE expression $ \text{SE} = ( 1 - \frac{\tau}{\tau_c} ) \log_2 ( 1+ \text{SINR} )$, increasing $\tau$ directly reduces the pre-log factor $( 1 - \frac{\tau}{\tau_c} )$, which in turn lowers the achievable SE.\\
Fig.~\ref{Ave. ve. APs} compares the impact of the AP configuration on the average uplink throughput of various AP selection schemes, for $U = 50$ and $\tau = 20$. The results reveal a clear upward trend in average uplink throughput with increasing AP count. This improvement can be attributed to the reduced path loss as APs are deployed more densely, along with a gradual decrease in inter-user interference. Across all configurations, the proposed DAPPA scheme consistently outperforms the competing methods, confirming its effectiveness and highlighting its robustness in leveraging additional APs to enhance system throughput.

\subsubsection{Results based on the power allocation with all APs}
\add{In this subsection, we compare the methods under a setting where only the power allocation scheme is optimized, while all APs are jointly considered. Fig.~\ref{CDF} illustrates the effect of power allocation in the absence of AP selection by showing the SE CDFs of the proposed DAPPA scheme and benchmark methods for two user densities, $U=40$ and $U=80$. In both scenarios, the proposed approach consistently shifts the CDF toward higher SE values, indicating improved user fairness and stronger overall performance. The advantage is most pronounced beyond the 95th percentile, where users typically suffer from moderate to poor channel conditions. As the user density increases to $U=80$, all schemes experience SE degradation due to higher interference and tighter resource constraints; nevertheless, the proposed method preserves a clear performance edge in the mid-percentile region and provides more reliable service to far users. These results demonstrate that the proposed power allocation strategy achieves a more favorable SE distribution than existing optimization-based and learning-based schemes.\\
\begin{figure}[h]
	\centering
\includegraphics[width=3.6in,height=2.1in]{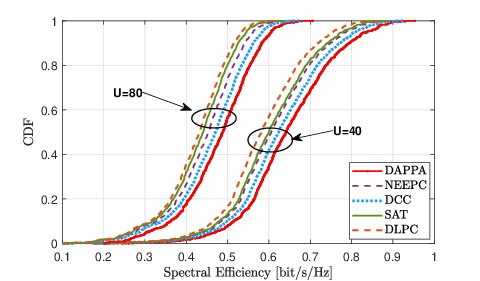}
	\caption{SE performance in the case with all AP and power allocation are applied to the system with $L = 100$ APs, $M=1$ antenna, $U = 40$ and $80$ users, and $\tau = 20$.}
 \label{CDF}
\end{figure} 
Fig.~\ref{Users all method} shows the average SE achieved by different pilot power allocation schemes as a function of the number of users \(U\), assuming that all APs serve all users. The results are shown for a system with \(L=100\) APs, a single antenna per AP (\(M=1\)), and pilot length \(\tau=20\). As the number of users increases, the average SE decreases for all schemes due to stronger pilot contamination and increased multi-user interference. Nevertheless, the proposed DAPPA scheme consistently outperforms the benchmark methods across all user densities. This performance gain highlights the effectiveness of the proposed pilot power allocation in mitigating interference and efficiently utilizing pilot resources. The gap between DAPPA and the other schemes becomes more pronounced as the network becomes more congested, demonstrating the robustness and scalability of the proposed approach in dense user scenarios.
\begin{figure}[h]
	\centering
\includegraphics[width=3.6in,height=2.1in]{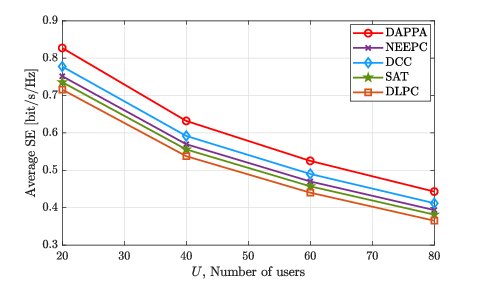}
	\caption{Performance of schemes with all APs and pilot power allocation for varying numbers of users, with $L = 100$ APs, $M = 1$ antenna, and $\tau = 20$.}
 \label{Users all method}
\end{figure} 
Fig.~\ref{tau all method} shows the average SE versus the pilot length $\tau$ for different AP selection and power allocation schemes, considering a system with $L=100$ APs, a single antenna per AP ($M=1$), and $U=40$ users. For all methods, the SE increases for small values of $\tau$, peaks around $\tau \approx 20$, and then gradually decreases as $\tau$ grows further. This behavior reflects the fundamental trade-off between pilot contamination and pilot overhead. When $\tau$ is small, the reuse of non-orthogonal pilots results in strong pilot contamination and poor channel estimation accuracy. Increasing $\tau$ improves pilot orthogonality and estimation quality, leading to higher SE. However, excessively large pilot lengths reduce the portion of the coherence block available for data transmission, which eventually dominates and causes the SE to decline.\\
Among all schemes, the proposed DAPPA method consistently achieves the highest SE across the entire range of $\tau$ values, owing to its effective joint AP selection and power allocation design. At $\tau = 20$, DAPPA reaches a peak SE of approximately $0.648$~bit/s/Hz, outperforming the second-best scheme, DCC, by about $5$--$8\%$. NEEPC follows closely behind, while SAT exhibits lower performance. DLPC performs the worst for all pilot lengths, indicating its limited capability in interference mitigation and power optimization.
Overall, these results emphasize the impact of the pilot length on system performance in CFmMIMO systems and demonstrate that the proposed approach is more resilient in sustaining higher SE across different pilot overhead regimes.
\begin{figure}[!]
	\centering
\includegraphics[width=3.6in,height=2.1in]{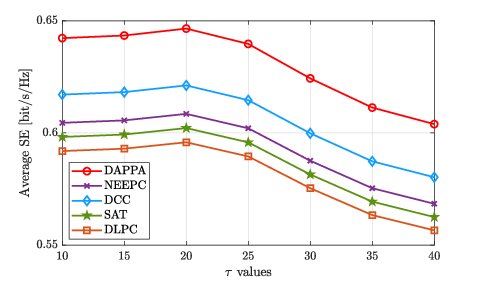}
	\caption{Average SE performance of different schemes with power allocation for varying $\tau$, with $L = 100$ APs, $M = 1$ antenna, and $U = 40$.}
 \label{tau all method}
\end{figure} 
Fig.~\ref{sumrate} evaluates the fairness performance of the proposed and benchmark schemes from the CDF of the user sum rate. From the CDF in Fig.~\ref{sumrate}, it can be seen that DAPPA achieves the highest probability of offering larger sum rates compared to all other methods. This gain is attributed to its iterative nature, which balances pilot and data power in every round, thereby mitigating pilot contamination and suppressing inter-user interference more effectively. DCC and NEEPC also yield improved distributions compared to conventional SAT and DLPC, but their tails shift to lower rates, indicating weaker guarantees for the worst-case users. 
\begin{figure}[!]
	\centering
\includegraphics[width=3.6in,height=2.1in]{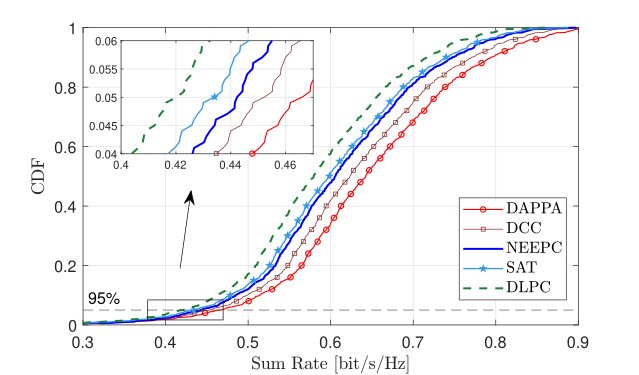}
	\caption{CDF of sum rate for $\tau=20$, with $L = 100$ APs, $M = 1$ antenna, and $U = 40$.}
 \label{sumrate}
\end{figure}}

\begin{figure}[!]
    \centering
    % First subfigure
    \subfloat[CDF of SE]{
        \includegraphics[width=0.51\textwidth]{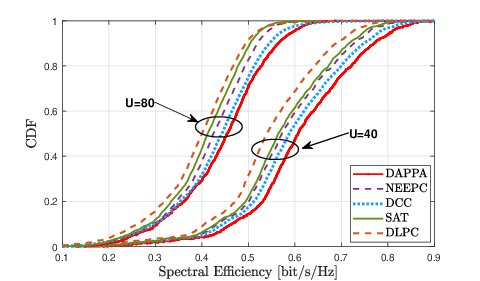}
        \label{CDF_SE_ALL}}  \vspace{-0.01cm} % Space between subfigures
    % Second subfigure
    \subfloat[95\% likely SE]{
        \includegraphics[width=0.51\textwidth]{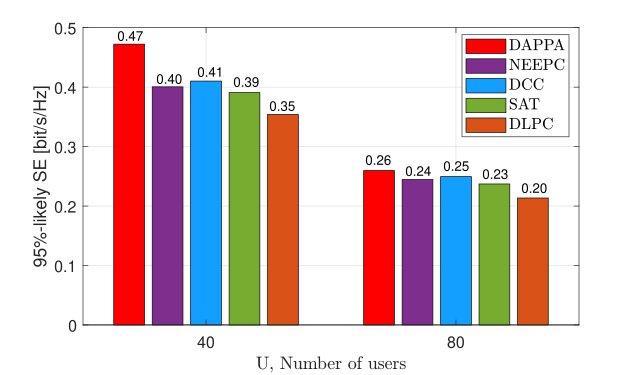}
        \label{fig:95percent}}
    \caption{SE performance in the case that the AP selection and power allocation are applied to the system with $L = 100$ APs, $M=1$ antenna, $U = 40$ and $80$ users, and $\tau = 20$.}
    \label{Bar 95 percent}
\end{figure}

\subsubsection{Results based on the AP selection and the power allocation approach}
In the second subsection, we evaluate the proposed method incorporating both AP selection and power allocation. Its performance is compared against several benchmark schemes: DCC\cite{bjornson2020scalable}, SAT\cite{mai2018pilot}, normalized estimation error-based power control (NEEPC)\cite{sarker2023pilot}, and deep learning-based power control (DLPC) \cite{zhang2021deep}.\\
\indent Fig.~\ref{Bar 95 percent} depicts the impact of AP selection and the power allocation. Fig.~\ref{CDF_SE_ALL} presents the CDF of the SE for the proposed DAPPA method and other benchmark schemes under two user densities, $U = 40$ and $U = 80$. For both cases, the CDF curves for the proposed DAPPA scheme indicate higher SE values for a larger proportion of users. This suggests that the proposed method improves fairness and ensures better average performance across users. The gain is especially visible for more than 80\% of the SE distribution, where users typically experience moderate and weak channel conditions. As the network becomes more congested with $U = 80$, the SE for all methods decreases due to increased interference and reduced resource availability; however, the proposed method maintains a noticeable advantage in the mid-percentile range and delivers more robust far user performance compared to the other approaches. These results highlight the proposed power allocation scheme’s ability to yield superior SE distribution, outperforming conventional optimization-based and learning-based power control strategies.\\
Fig.~\ref{fig:95percent} presents the 95\%-likely SE for the proposed scheme and other benchmark methods using $L = 100$ APs, $M = 1$ antenna, $\tau = 20$, and two different user counts, $U = 40$ and $U = 80$. For $U = 40$, the proposed DAPPA scheme outperforms the next-best method DCC by approximately $14.6\%$, while achieving gains of about $17.5\%$ over NEEPC, $20.5\%$ over SAT, and $34.3\%$ over DLPC. When the user count increases to $U = 80$, all schemes experience a significant SE reduction due to increased interference and reduced time-frequency resources; however, the proposed method still maintains an advantage of around $4\%$ over DCC, $8.3\%$ over NEEPC, $13.0\%$ over SAT, and $30.0\%$ over DLPC. These results demonstrate that the proposed approach consistently achieves a higher reliability-oriented SE, with performance gains sustained across both moderate and dense network conditions.
\begin{figure}[!]
	\centering
\includegraphics[width=3.6in,height=2.1in]{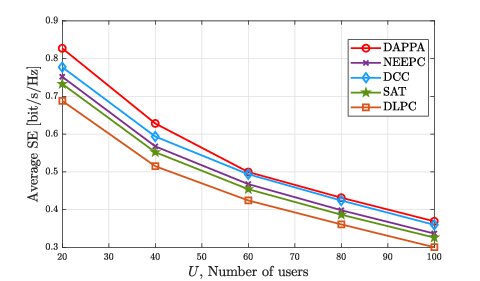}
	\caption{Performance of AP selection and power allocation schemes for varying numbers of users, with $L = 100$ APs, $M = 1$ antenna, and $\tau = 20$.}
 \label{Users all method}
\end{figure} 

\indent Fig.~\ref{Users all method} illustrates the fairness performance evaluation with the number of users $U$ for the setup of $L = 100$ APs, $M = 1$ antennas, and $\tau = 20$. In all cases, the SE decreases monotonically as the number of users increases, which is expected since more users lead to increased inter-user interference and reduced per-user resource allocation.
The proposed DAPPA scheme consistently outperforms all benchmark schemes across the entire user range. At $U = 20$, it achieves the highest SE of approximately $0.83$~bit/s/Hz, which is about $10\%$-$15\%$ higher than the best-performing baseline DCC. As $U$ increases, the SE advantage of the proposed method persists, although the gap gradually narrows, reaching around $0.36$~bit/s/Hz at $U = 100$. Among the baselines, DCC shows the closest performance to the proposed method, followed by NEEPC and SAT. DLPC consistently yields the lowest SE, indicating that its power control strategy is less effective in managing interference under increasing user density. Notably, the performance difference between NEEPC and SAT is relatively small, especially in high-load scenarios ($U \geq 80$), suggesting similar interference-limited behavior. Overall, it can be seen that the AP selection and power allocation strategies of the proposed method provide a clear performance benefit in both light and heavy load conditions, with the most significant gains observed in the low-to-moderate user regime.\\
\indent Fig.~\ref{tau all method} presents the average SE as a function of the pilot length $\tau$ using the AP selection and power allocation schemes, with $L = 100$ APs, $M = 1$ antenna, and $U = 40$ users. Across all schemes, the SE initially increases with $\tau$, reaches a peak around $\tau \approx 15$, and then gradually decreases as $\tau$ continues to grow. This trend can be explained as follows: when $\tau$ is small, pilot contamination is severe due to non-orthogonal pilot reuse, leading to inaccurate channel estimation. Increasing $\tau$ improves pilot orthogonality and channel estimation quality, thus enhancing SE. However, beyond a certain point, a larger $\tau$ reduces the fraction of the coherence block available for data transmission, which outweighs the benefit of improved estimation accuracy, causing the SE to decline.\\
The proposed DAPPA scheme consistently achieves the highest SE across all $\tau$ values, benefiting from its AP selection and power allocation strategies. At $\tau = 15$, it attains the maximum SE of approximately $0.635$~bit/s/Hz, which is about $3$-$5\%$ higher than the second-best method DCC. NEEPC forms the next performance tier. SAT follows, while DLPC performs the worst for all $\tau$, indicating limited effectiveness in mitigating interference and optimizing power usage.
This behavior highlights the trade-off between pilot overhead and channel estimation quality in CFmMIMO, and shows that the proposed method is more robust in maintaining higher SE over the entire range of $\tau$ values.
\begin{figure}[!]
	\centering
\includegraphics[width=3.6in,height=2.1in]{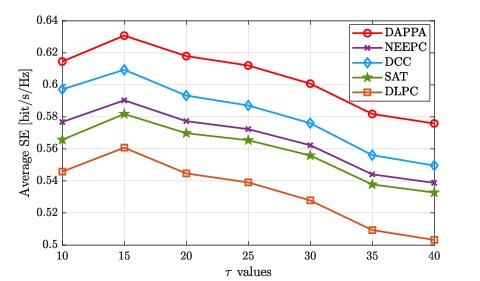}
	\caption{Average SE performance of different AP selection and power allocation schemes for varying $\tau$, with $L = 100$ APs, $M = 1$ antenna, and $U = 40$.}
 \label{tau all method}
\end{figure}

\section{Conclusion}
This paper proposed a dynamic AP selection and pilot power allocation (DAPPA) framework for uplink CFmMIMO systems. Using hierarchical correlation-based clustering, users are associated with APs offering strong channel gains and low correlation, ensuring reliable connectivity, effective interference mitigation, and flexible cluster adjustment without full network reclustering. A user-capacity constraint per AP was introduced to improve scalability, while pilot power allocation was formulated as a WSRM problem and solved iteratively via quadratic transform to enhance SINR and guarantee fairness. Numerical results showed that DAPPA achieves higher spectral efficiency, faster convergence, and robust performance in dense multi-user scenarios compared to benchmark schemes, making it a scalable solution for next-generation CFmMIMO networks.

\ifCLASSOPTIONcaptionsoff
  \newpage
\fi

%  \appendices

\bibliographystyle{IEEEtran}
\bibliography{refs}

\end{document}